\documentclass{optica-article}

\journal{opticajournal} % for journals or Optica Open

\articletype{Research Article}

\usepackage{lineno}
\begin{document}

\title{In vacuum metasurface for optical microtrap array}

\medskip

\author{Donghao Li,\authormark{1,2,$\dag$} Qiming Liao,\authormark{3,$\dag$} Beining Xu,\authormark{1} Thomas Zentgraf,\authormark{4} Emmanuel Narvaez Castaneda,\authormark{4} Yaoting Zhou,\authormark{1} Keyu Qin,\authormark{1} Zhongxiao Xu,\authormark{1,2,*} Heng Shen,\authormark{1,2,*} and Lingling Huang\authormark{3,*}}

\address{\authormark{1}State Key Laboratory of Quantum Optics Technologies and Devices, Shanxi University, Taiyuan, 030006, China\\
\authormark{2}Collaborative Innovation Center of Extreme Optics,Shanxi University, Taiyuan, 030006, China\\
\authormark{3}School of Optics and Photonics, Beijing Engineering Research Center of Mixed Reality and Advanced Display, Beijing Institute of Technology, Beijing 100081, China\\
\authormark{4}Department of Physics, Paderborn University, Paderborn, Germany\\
\authormark{\dag}The authors contributed equally to this work.\\
\authormark{*}xuzhongxiao@sxu.edu.cn, hengshen@sxu.edu.cn, huanglingling@bit.edu.cn}
\medskip

\begin{abstract*} 
Optical tweezer arrays of laser-cooled and individual controlled particles have revolutionized the atomic, molecular and optical physics, and they afford exquisite capabilities for applications in quantum simulation of many-body physics, quantum computation and quantum sensing. Underlying this development is the technical maturity of generating scalable optical beams, enabled by active components and high numerical aperture objective. However, such a complex combination of bulk optics outside the vacuum chamber is very sensitive to any vibration and drift. Here we demonstrate the generation of 3$\times$3 static tweezer array with a single chip-scale multifunctional metasurface element in vacuum, replacing the meter-long free space optics. Fluorescence counts on the camera validates the successfully trapping of the atomic ensemble array. Further, we discuss the strategy to achieve low scattering and crosstalk, where a metasurface design featuring dual-wavelength independent control is included. Our results, together with other recent development in integrated photonics for cold atoms, could pave the way for compact and portable quantum sensors and simulators in platforms of neutral atom arrays. 
\end{abstract*}

%%%%%%%%%%%%%%%%%%%%%%%%%%  body  %%%%%%%%%%%%%%%%%%%%%%%%%%
\section{Introduction}
 Assembling the individually controlled quantum objects lies in the heart of quantum physics. Recent decade has witnessed great developments in the experimental control over arrays of individually neutral atoms \cite{endres_atom-by-atom_2016,barredo_atom-by-atom_2016,kim_situ_2016,browaeys_many-body_2020,ZhanMS_twospcies_2020}, emerging as a promising platform for quantum computation \cite{bluvstein_quantum_2022,evered_high-fidelity_2023,ma_high-fidelity_2023,graham_multi-qubit_2022,bluvstein_logical_2024}, simulation of many-body physics \cite{bernien_probing_2017,semeghini_probing_2021,ebadi_quantum_2022,scholl_quantum_2021,keesling_quantum_2019,bluvstein_controlling_2021,chen_continuous_2023} and quantum sensing. In this configuration, single neutral atoms or atomic ensembles are initially trapped and assembled in arrays of tightly focused laser beams, so-called static optical tweezers, while fast moving and single addressing are enabled by dynamical optical tweezers. To generate such static microtrap arrays, one typically leverages active components, such as spatial-light-modulator (SLM) \cite{barredo_atom-by-atom_2016,ebadiQuantumPhasesMatter2021}and digital-micromirror devices (DMD)\cite{wangPreparationHundredsMicroscopic2020}, or microlens arrays\cite{tanakaKaleidoscopicPatterningMicro2017,schaffnerArraysIndividuallyControllable2020} in combination with conventional high-numerical-aperture (NA) objectives. A meter-long optical path defined by these elaborated elements is however susceptible to platform vibrations and drift, degrading the beam-pointing stability and system performance. 

In practice, integrated photonics has recently benefited atomic, molecular and optical (AMO) experiments through miniaturizing the beam delivery components and improving the reliability. For instance, silicon nitride ($\text{Si}_3\text{N}_4$) waveguides and grating couplers were exploited to deliver the magneto-optical trap (MOT) beams in cold atoms \cite{CLZ_gratingtrap_2022,isichenko_photonic_2023} and all wavelengths of light necessary to control the ion qubit \cite{ionCa_integrated_2020,niffenegger_integrated_2020}. Yet multifunctional and scalable optical components is highly demanded, especially in applications outside the laboratory such as mobile quantum sensors.  

Metasurfaces, two-dimensional structures composed of optically thin arrays of subwavelength dielectric or metallic scatters, offer the opportunity to fully control light\cite{hePluggableMultitaskDiffractive2024,xuQuasicrystalMetasurfaceDual2024,yinMultiDimensionalMultiplexed2024,heMetaAttentionNetwork2024,shaoMultispectralImagingMetasurface2024,suPlanarChiralMetasurface2024,Zhan_OE_2024}. These structures exhibit the multidimensional control ability, for example, polarization modulation, as well as simultaneously modulation of various fundamental properties of light, which greatly surpass the capabilities of diffractive optical elements. 

Conventional optical tweezer arrays are typically generated using bulky devices such as spatial light modulators (SLMs), acousto-optic deflectors (AODs), and digital micromirror devices (DMDs). While these established approaches enable dynamic control of optical potentials, they present several inherent limitations: the macroscopic dimensions of these components result in complex optical setups requiring precise alignment, and their operation demands sophisticated electronic control systems. Moreover, the generation of high-quality optical trap arrays usually need additional optical elements to compensate the inherent aberrations introduced by conventional devices, further complicating system integration and stability. In contrast, emerging metasurface technology provides a compact platform for advanced optical manipulation.  By precisely engineering nanostructures of metasurface, tailored phase modulation of optical fields can be achieved to enable sophisticated wavefront control. The planar geometry design simplifies optical path configuration while significantly reducing alignment complexity. With its high numerical aperture, it enables direct atomic trapping at the focal plane without additional optical components. These advantages make metasurface particularly promising for developing next-generation quantum simulation platforms. Pioneering works have demonstrated their capabilities of addressing the specific tasks aided with bulk free-space optic components, such as high-efficiency MOT beam delivery \cite{zhu_dielectric_2020} and high-NA optical objective \cite{hsu_single-atom_2022}. However, multifunctional photonic elements are feasible by manipulating the spatial patterns of nanostructures, which enables the beam steering, local control of optical polarization and enhancement of the emission and detection.

Here, we report the demonstration of optical metasurfaces enabled laser dipole trapping of $^{87}$Rb atomic arrays. To our best knowledge, it is the first time to generate the static optical tweezer array using a single chip-scale photonic element in vacuum, which replaces the meter-long free-space optics including active component (e.g. SLM and AOD) and high-NA objectives. In the experiment, a collimated laser beam with the waist about 1 mm illuminates the metasurface chip in vacuum, producing a 3$\times$3 optical tweezer array to trap atomic ensembles. Fluorescence counts on the charge-coupled-device (CCD) camera indicates our successful trapping. We further discuss the strategy to achieve low scattering and crosstalk, where a metasurface design featuring wavelength-selective control is presented.

\section{Metasurface design and characterization}
\begin{figure}[htbp]
    \centering
    \includegraphics[width=0.95\linewidth]{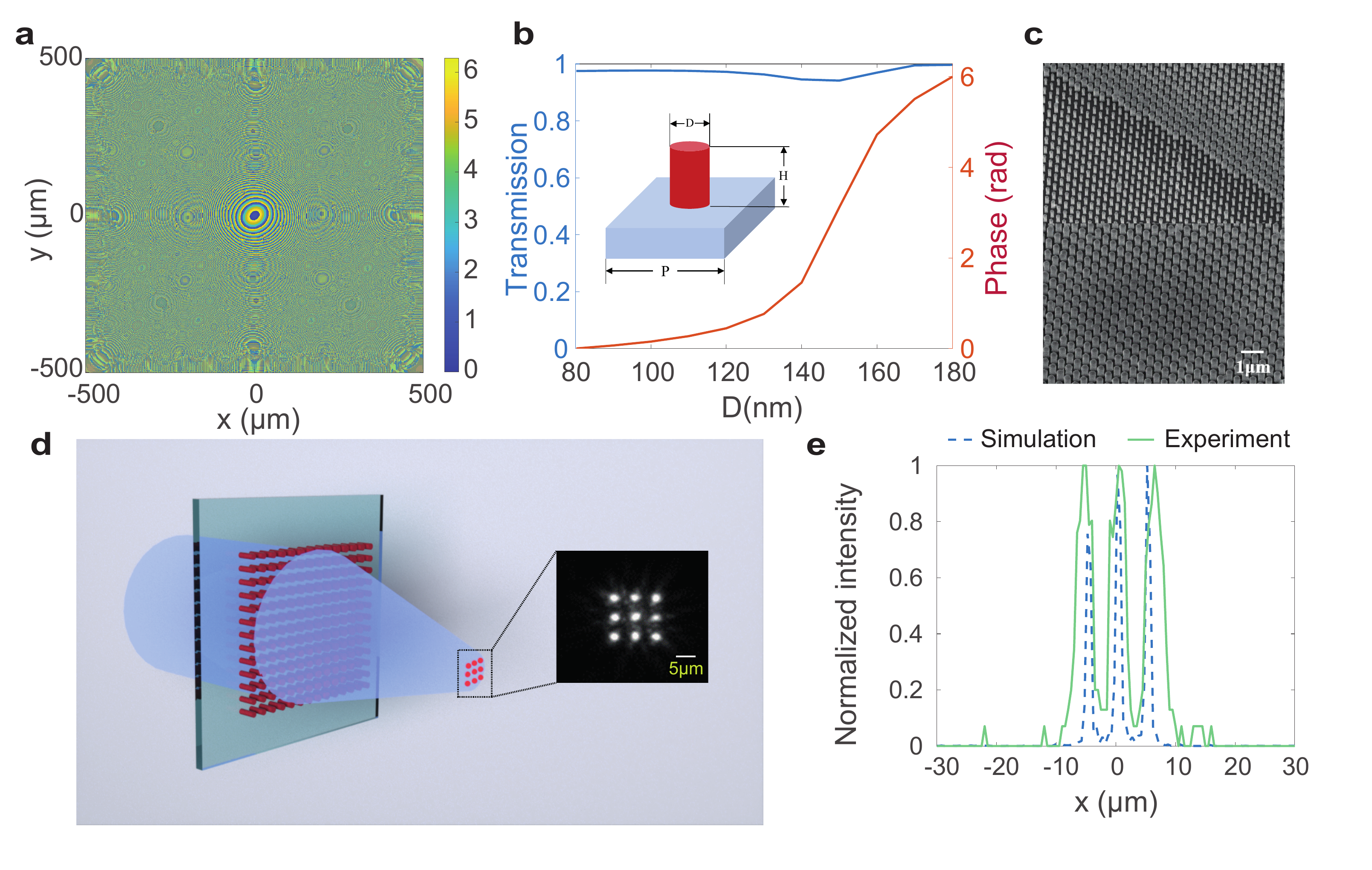}
    \caption{\textbf{Metasurface design and characterization.} \textbf{a,} The phase profile of the metasurface generated using the optimized GS algorithm. \textbf{b,} Transmission coefficients and phase obtained by sweeping the diameter of a nanopillar within a unit cell. \textbf{c,} Top-view and side-view SEM images. \textbf{d,} The schematic diagram of the focal spot array focusing under 852 nm incident light and the experimentally captured image of the focal plane. \textbf{e,} The comparison of the energy distribution of the middle row of focal spots between the experimental results and the simulation results. }
    \label{Fig1}
\end{figure}

To create a 3 $\times$ 3 focal spots array with a designed focal length of 1 mm and a spacing between each spot of 5 $\mu$m, we employ a modified Gerchberg-Saxton (GS) algorithm to generate the desired phase-only hologram. As a classical and efficient holography algorithm, the GS algorithm has the advantages of high adaptability and reconstruction accuracy. To enable the flexible design of the focal plane position, we propose an optimization scheme that introduces angular spectrum diffraction as the propagation kernel in the GS algorithm, transforming the design space from the Fourier plane to the Fresnel domain. By replacing the fast Fourier transform with angular spectrum diffraction and specifying the diffraction distance as the focal length, we obtained the phase profile as shown in Fig. \ref{Fig1}a that can achieve a high-quality multiple focal spots array. Due to the introduction of diffraction theory into calculation, this method can produce the focal spots array with more uniform energy distribution and higher signal-to-noise ratio compared with the traditional design method, which only simply superimposing multiple lens phases. See Supplementary Material Note 1. 
The designed metalens has a diameter of 1 mm and a focal length of 1 mm, resulting in a numerical aperture (NA) of 0.447. This NA is sufficient for effective optical trapping of cold atoms, enabling sub$-$micron beam waists suitable for single atom tweezer applications.

    Then we design and realize a polarization-independent metasurface using amorphous silicon nanopillars patterned on a fused silica substrate which can operate at $\lambda$ = 840 nm, the working wavelength for trapping rubidium atoms. We calculate the transmission coefficient and phase responses of meta-atoms with different diameters at 840 nm using the rigorous coupled-wave analysis (RCWA) method. To cover the phase shifts from 0 to $2\pi$ with high efficiency, the period and height of the nanopillars are chosen as 500 and 600 nm respectively and the diameter ranges from 80 nm to 180 nm. As shown in Fig.\ref{Fig1}b, the selected structures have transmission coefficients exceeding 0.9 and can cover the phase range from 0 to $2\pi$. By engineering the nanopillars, the desired phase profile can be converted into a diverse diameter distribution across the metasurface. The fabricated metasurface is composed of 2002 $\times$ 2002 nanopillars using electron beam lithography and reactive ion etching, and the corresponding scanning electron microscopy images with side and top views are shown in Fig. \ref{Fig1}c.

    Subsequently, we test the focusing performance of the fabricated sample under incident light with a wavelength of 840 nm. At a distance of 1.01 mm from the metasurface, we obtain the clearest image of a 3 $\times$ 3 focal spots array, as shown in Fig. \ref{Fig1}d, with the average distance between the focal spots characterized as 5.2 $\mu$m and an average full width at half maximum (FWHM) of 3.03 $\mu$m. For comparison, we simulate the energy distribution of the focal points in the middle row by using angular spectrum diffraction method. It can be observed in Fig. \ref{Fig1}e that the experimental results exhibit a more uniform energy distribution, which may be due to phase deviations from the design values caused by fabrication errors, resulting in partial dispersion of the energy that should have been entirely focused at the focal spots. This also leads to a larger FWHM in the experimental results and a 0.2 $\mu$m difference in the spacing between focal spots compared to the design value of 5 $\mu$m. We also characterized the energy distribution of the focal spots in other rows and columns, as well as the energy distribution in the x-z plane.  Detailed information can be found in Supplementary Material Note 2. The experimental results demonstrate that the metasurface can generate a uniform focal spots array, providing feasibility for subsequent optical tweezer experiments.

\section{Optical metasurface-enabled trapping of atom array}
\begin{figure}
    \centering
    \includegraphics[width=0.9\linewidth]{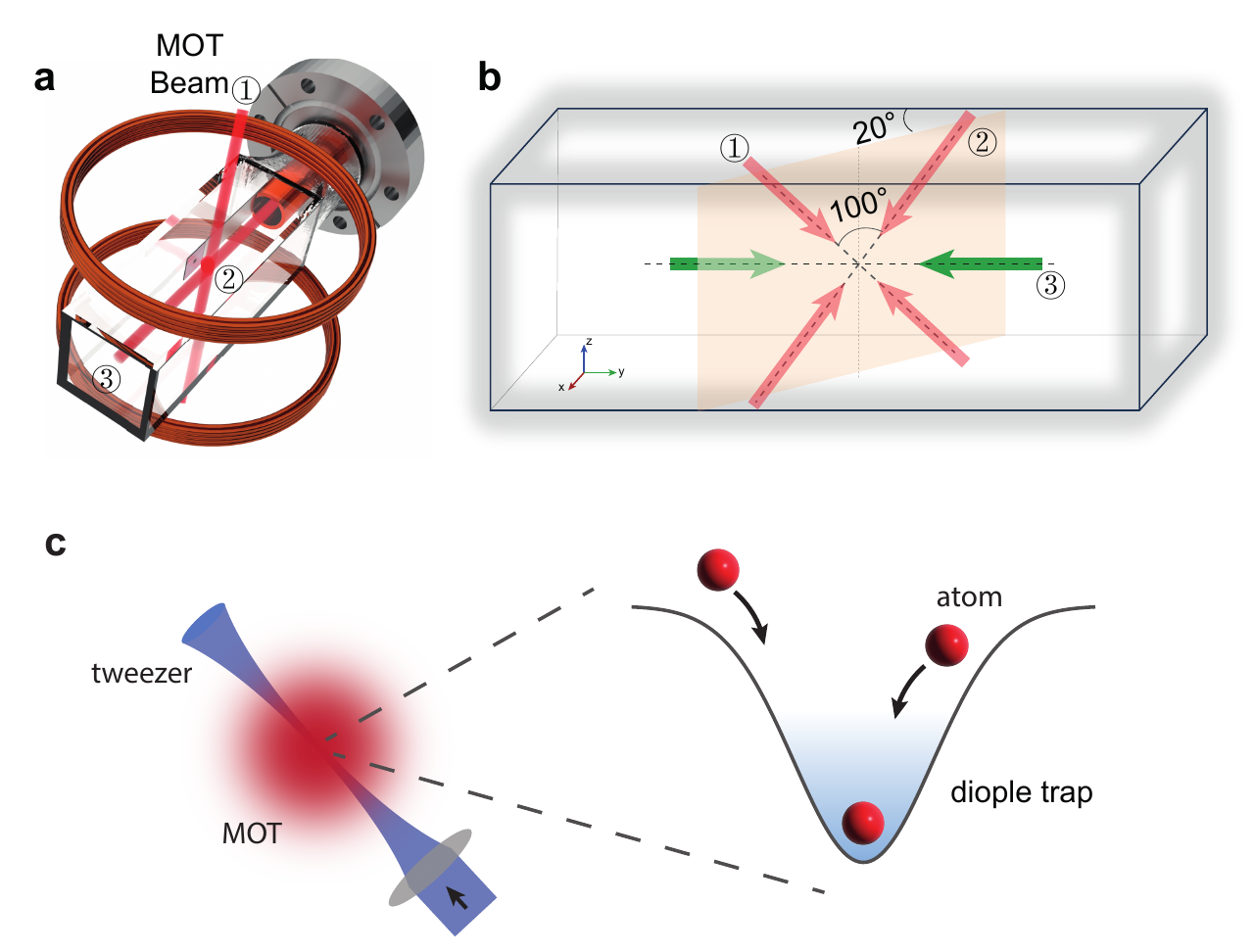}
    \caption{Overview of experiment setup \textbf{a,} MOT setup. Three pairs of MOT beams are overlapped at the zero-magnetic-field location generated by the anti-Helmholtz coils.  \textbf{b,} MOT beams geometry. \textbf{c,} Trapping process of the diople trap. Rb atoms are loaded from a magneto-optical-trap (MOT) into an optical dipole trap that composed of a focused, far-off-reasonance beam.}
    \label{Fig5}

\end{figure}

 The microstructure (1 mm $\times$ 1 mm) with nanopillars is located at the center of substrate (10 mm $\times$ 10 mm).  The metasurface sample is bonded to a glass plate and then fixed inside our vacuum system. Due to the geometric restriction of metasurface, one pair of 3D MOT beams is parallel to the metasurface along the y direction, and the other two pairs intersect with an angle of $\sim$ 100 $^{\circ}$ inside a plane. The angle between this plane and y-z plane is about 20$^{\circ}$, as illustrated in Fig. \ref{Fig5}b. This 3D MOT configuration is a trade-off between atomic number and metasurface structure. 

\begin{figure}
    \centering
    \includegraphics[width=0.95\linewidth]{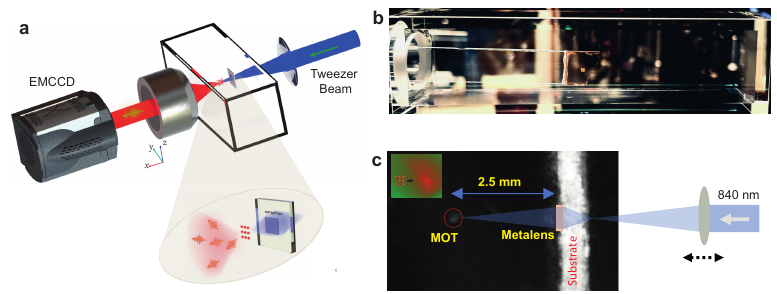}
    \caption{\textbf{Metasurface trapping setup.} \textbf{a,} Overview of trapping and imaging. A laser beam at $\lambda$=840 nm passes through the metasurface and then a 3 $\times$ 3 tweezer array is created. The atomic fluorescence at $\lambda$=780 nm is collected by a objective lens and imaged on a EMCCD camera. We use the forward collection method that several filters are used before the camera (not shown here). \textbf{b,} Photograph of the metasurface and its mounting structure. The metasurface is first glued to a glass plate, which is fixed on the flange of glass cell. \textbf{c,} Schematics of the laser illumination. A collimated beam first passes through an adjustable lens, resulting in a focus before the metasurface sample. This diverging beam illuminates the metasurface and consequently produces the trap tweezers. By moving the adjustable lens, we can shift the focus position to ensure the overlap with 3D MOT.}

    \label{Fig2}
\end{figure}

In optical trapping process, both metasurfaces and traditional objective lenses rely on the principle of generating tightly focused laser beams with high NA to create strong intensity gradients, thereby producing dipole forces to trap particles \cite{grimm1999opticaldipoletrapsneutral}. Experimentally, the tweezer is immersed in atomic ensemble, where trapping is accomplished through the dipole force generated by tightly focused Gaussian beam (Fig. \ref{Fig5}c).
 A diagram of the atom trapping through optical metasurface is illustrated in Fig. \ref{Fig2}a. The metasurface sample is mounted inside a glass vacuum cell (24 mm$\times$28 mm$\times$100 mm) with an antireflection coating on the outside (Fig. \ref{Fig2}b), and a high vacuum pressure of $1\times 10^{-6}$ Pa is maintained with an ion pump. The experiment starts with the three-dimensional MOT, where a cloud of rubidium atoms are cooled in the cross section of three retro-reflected MOT laser beams. Such 1 mm-diameter MOT beams with a total power of 500 $\mu$W are red detuned for the $^{87}$Rb D2 $F$=2 to $F'$=3 transition by 19 MHz, and are spatially overlapped with repump beams at a power of 50 $\mu$W resonant with D1 $F$=1 to $F'$=1 transition. The axial magnetic field gradient is 16 G/cm through the trap center. This MOT loading lasts for 200 ms, and the absorption imaging time-of-flight (TOF) measurement give the size of $1.2\times 10^6$ atoms and temperature of 160 $\mu$K.
    
   Subsequently, a optical tweezer array is switched on, overlapping with the 3D MOT. To produce this tweezer array, a laser beam is delivered from a continuous wave Ti:S single frequency laser at $\lambda$ = 840 nm, and spatially filtered by a polarization-maintaining single-mode fiber terminated with zoom fiber collimator (Thorlabs ZC618APC-B). The normal incidence of this beam at the metasureface sample induces a 3 $\times$ 3 beam array through the optical diffraction. Notice that the optical diffraction on the sample is polarization insensitive. And the trapped atom array is imaged in the electron-multiplying charge-coupled device (EMCCD, Andor iXon Ultra) by collecting the fluorescence via an objective (Mitutoyo G Plan Apo 50$\times$), as shown in Fig. \ref{Fig2}a. Since we collect the forward scattering fluorescence at 780 nm, three narrow-band filters (Lier - opto, LO78010254ZD) are used in front of the camera to block the transmitted trap beam at 840 nm.

\section{Imaging the atomic array}

In practice, the designed focus plane is not employed when accounting for the MOT configuration, and the associated reason is as follows. No wavelength-selection in the design of metasurface sample causes the same diffraction of MOT beams at 780 nm as that of trap beam at 840 nm, and all MOT beams thus have to avoid the incidence on this optical element. We consider all the factors, such as 3D MOT geometry, background scattering from the sample holder and the probe beam delivery for trapped atom imaging, and finally decide to shift the focus plane from 1 mm to 2.5 mm with respect to the metasurface by using a lens (Fig. \ref{Fig2}c). Experimental results show the average distance of focal spots is about 13 $\mu$m, and a $1/e^2$ waist each spot is 5.5 $\mu$m. More details can be found in Supplementary Material Note 3. 
\begin{figure*}[htbp]
    \centering
    \includegraphics[width=0.9\textwidth]{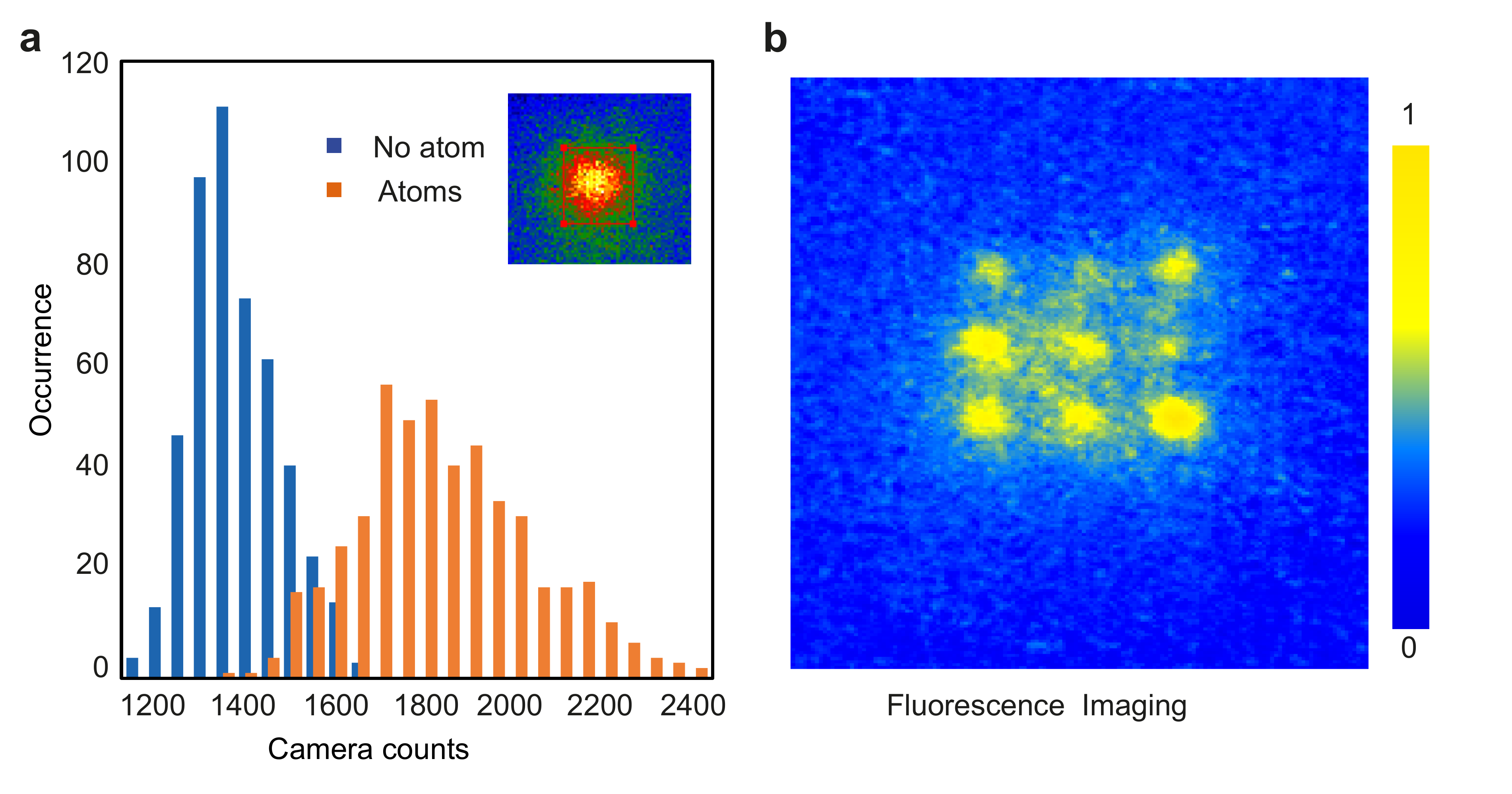}
    \caption{\textbf{Imaging the atomic array in optical microtraps.} \textbf{a,} Fluorescence counts distribution.The blue bars represent the background counts while the orange bars represent the atomic fluorescence counts. \textbf{b,} Fluorescence image of the atomic array in the metasurface enabled optical tweezers.}
    \label{Fig3}
\end{figure*}

In Fig. \ref{Fig3}a, we plot a histogram of the fluorescence counts registered on the EMCCD camera (400 ms) versus the occurrence from the red square region marked in the insert (a single microtrap in the tweezer array, 25 pixels), where 500 loading interactions are averaged. The lower-count histogram peak in blue represents the background signal with no atom, while the higher-count peak in orange corresponds to the signal with atoms. However, the collision blockade effect is prevented \cite{schlosser_sub-poissonian_2001,schlosser_collisional_2002} by the larger beam size (5.5 $\mu$m) induced by the focus plane shift with respect to the original design (1 $\mu$m). Consequently, more than one atoms are loaded in each of the optical tweezer. An averaged fluorescence image of 3$\times$3 atomic array is exhibited in \ref{Fig3}b. The inhomogeneity is attributed to the inhomogeneous intensity distribution of trap beam array. And this is a consequence of the focus plane shift as mentioned above. 
\section{Discussion}

    In summary, we report the first demonstration of static optical tweezer array with a single chip-scale metasurface sample in vacuum. A 3$\times$3 array of atomic ensembles is confined in such optical microtraps, which is verified through the fluorescence counts on the CCD camera. Our work sets an example of integrated photonics facilitate cold atom system. In combination with integrated optic 3D MOT \cite{isichenko_photonic_2023,zhu_dielectric_2020}, it offers the possibility for portable quantum sensors with neutral atom array platform in unprecedented ways \cite{schaffner_sensing_2024,Kufman_squeezing_2023,Endres_sMultiensemble_2024}. Intriguingly, our work sheds light on an entirely new architecture of optical metasurface enabled quantum metasurface, where the nano-scatters are replaced with atoms \cite{Yelin_qmetasurface_2020}.
     
     \begin{figure*}[htbp]
    \centering
    \includegraphics[width=0.9\textwidth]{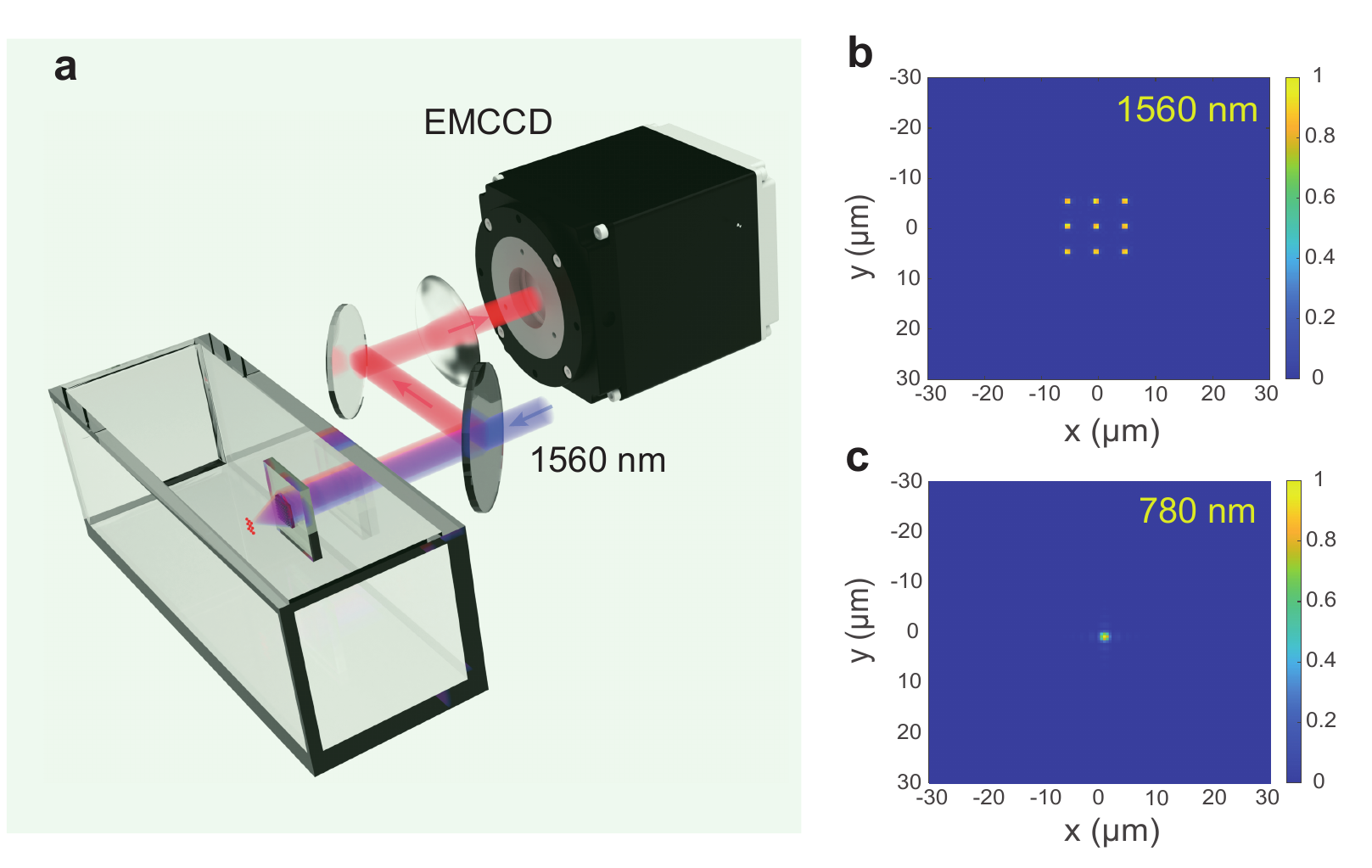}
    \caption{\textbf{New generation of dual-wavelength mtasurface design.} \textbf{a,} Blueprint of the new experimental setup. The normal incidence of a trap beam at 1560 nm on the metasurface sample forms a 3 $\times$ 3 tweezer array. While the backward fluorescence is collected by the metasurface and imaged on the EMCCD. \textbf{b,} Focus spots formed by a collimated 1560 nm input beam. \textbf{c.} Focus spot formed by a collimated 780 nm input beam. For 1560 nm trap beam, metasurface will generate a 3 $\times$ 3 tweezer array. For 780 nm atomic fluorescence, metasurface acts as a high NA collective lens. }
    \label{Fig4}
\end{figure*}

  Despite the successful trapping, two key challenges for getting the best performance exist. On one hand, the designed metasurface does not perform distinct functions for the incident light and atomic excitation fluorescence wavelengths. Therefore, a transmission-based system is required for fluorescence collection, which undoubtedly increases the experimental difficulty in background scattering. To address this issue, an improved scheme is proposed, a polarization-independent dual-wavelength version. Consider the dual-wavelength high power laser from Precilasers (780 nm and 1560 nm) available in the lab, this scheme utilizes the  strong optical field manipulation capabilities of metasurface to achieve different functions at the incident light at 1560 nm and the fluorescence light at 780 nm, respectively. When 1560 nm light is incident on the metasurface, as shown in Fig. \ref{Fig4}b, it generates a 3$\times$3 focal spots array similar to the existing sample. For the fluorescence at 780 nm, it only performs the focusing function as shown in Fig. \ref{Fig4}c. This allows for the backward collection of the excited fluorescence (Fig. \ref{Fig4}a), thereby simplifying the overall experimental complexity and improving the collection efficiency.
 
To achieve polarization-independent dual-wavelength independent control in the new design, various types of meta-atom structures will be combined, including cross-shaped pillars, cylindrical pillars, and ring-shaped pillars, to meet the design requirements. The phase response of these scanned structures needs to simultaneously cover a phase space of 0-2$\pi$ at both 1560 nm and 780 nm wavelengths. Subsequently, these structures will be encoded into the pre-designed dual-wavelength phase profiles to realize a multi-focal array at 1560 nm and a focusing function at 780 nm.

On the other hand, shifting the focus plane is required as a result of the 3D MOT configuration in the presented work, leading to the enlarged tweezer beams that trap more than one atoms. In the updated design aforementioned, focal length is 4 mm and the beam waist is 1 $\mu$m, where the collision blockade condition is promised and the effect of stray light from the substrate of metasurface sample is reduced. Further improvement in the loading rate can be achieved through the polarization-gradient cooling (PGC), cooling the atomic temperature to 40 $\mu$K (See supplementary materials for preliminary results). This approach can directly applied to the generation of a large scale atom array. 

Overall, our research demonstrates a static optical tweezer array in vacuum based on a single chip$-$scale metasurface, successfully trapping a 3$\times$3 array of atomic ensembles and showcasing a promising strategy for integrating photonics with cold atom systems. The introduction of a polarization$-$independent dual-wavelength metasurface significantly enhances fluorescence collection efficiency while reducing experimental complexity. This approach paves the way for scalable neutral atom platforms and offers a compelling route toward the realization of next$-$generation quantum metasurfaces.

\begin{backmatter}
\bmsection{Funding}
This work was supported by the National Key R\&D Program of China (2020YFA0309400), National Natual Science Foundation of China (12222409, 12204289) and Beijing Natural Science Foundation (JQ24028). H. S. acknowledges financial support from the Royal Society Newton International Fellowship Alumni follow-on funding (AL201024) of the United Kingdom.

\bmsection{Acknowledgment}
H.S.and L.H. conceived the idea. Q.L. T.Z.,E.N.C.and L.H. designed and fabricated the metasurface sample. D.L, B.X, Y.Z., K.Q. and Z.X. performed the laser cooling and trapping of atoms. D.L., Z.X.and H.S. contributed to the data analysis.  D.L., Q.L., L.H. and H.S. wrote the manuscript with contribution from all authors.

\bmsection{Disclosures}
The authors declare no conflicts of interests.

\bmsection{Data Availability Statement}
The data are available from the corresponding author on reasonable request.

\bmsection{Supplemental document}
See \href{Supplementary.pdf}{Supplementary Material} for supporting content.

\end{backmatter}

\bibliography{meta_citation}

\end{document}